\documentstyle[aps,epsf,multicol]{revtex} 
\begin{document}
\title{ 
Critical numbers of attractive Bose-condensed atoms in asymmetric traps
}
\author{A. Gammal$^{(a)}$, Lauro Tomio$^{(b)}$ and T.Frederico$^{(c)}$}
\address{ 
$^{(a)}$ Instituto de F\'{\i}sica, Universidade de S\~ao Paulo,
05315-970 S\~{a}o Paulo, Brazil \\
$^{(b)}$ Instituto de F\'\i sica Te\'orica, Universidade Estadual Paulista,
01405-900 S\~{a}o Paulo, Brazil \\
$^{(c)}$Departamento de F\'{\i}sica, Instituto Tecnol\'ogico da
Aeron\'autica, CTA,
12228-900 S\~ao Jos\'e dos Campos,
SP, Brazil }
\date{\today}
\maketitle
\begin{abstract}
The recent Bose-Einstein condensation of ultracold atoms with attractive 
interactions led us to consider the novel possibility to probe the 
stability of its ground state in arbitrary three-dimensional harmonic 
traps. We performed a quantitative analysis of the critical number of 
atoms through a full numerical solution of the mean field 
Gross-Pitaevskii equation.
Characteristic limits are obtained for reductions from three to two and 
one dimensions, in perfect cylindrical symmetries as well as in deformed 
ones.
\end{abstract}

\begin{multicols}{2}
The predicted collapsing behavior of the condensed systems with attractive 
two-body atomic interactions~\cite{rup95}, first observed in experiments 
with $^7$Li~\cite{hulet95}, was recently tested in experiments with 
$^{85}$Rb~\cite{roberts}. In the experiments described in Ref.~\cite{roberts}, 
and more recently in \cite{strecker}, by means of Feshbach resonance 
techniques~\cite{avaria}, the two-body interaction was tuned from positive 
to negative values.
Besides the fact that the experimental results qualitatively agree with the 
theory, and confirm results of previous variational 
treatments~\cite{sackett,baym,shuryak,stoof,ueda,wadati}, they also show a 
consistent quantitative deviation of about 20\% from the mean-field predicted 
critical number of atoms, $N_c$~\cite{rup95}. The asymmetry of the trap was 
shown in Ref.~\cite{GFTbr} to be responsible for about 4\% of the observed 
deviation. 

In this respect, it is relevant to obtain precise and reliable 
numerical results for the 
mean-field calculations, in order to probe their 
consistency and possible limitations. The actual experimental atomic traps are 
in general harmonic and non-symmetric.
Extreme asymmetric traps have been recently employed in 
experimental investigations with condensates constrained to quasi-one 
(1D)~\cite{strecker,schreck} or quasi-two dimensions (2D)~\cite{gorlitz}, 
exploring theoretical analysis considered by several 
authors~\cite{perez97,perez98,carr,baym,bhaduri,adhikari}.
A non symmetric three-dimensional (3D) trap is reported in 
Ref.~\cite{hensinger}, with the frequencies given by 
$2\omega_1=\sqrt{2}\omega_2=\omega_3=2\pi\times 33$Hz. 

Considering the general non-symmetric traps that have been employed,
the accuracy of the comparison between experiments and the results of 
mean-field approximation relies in precise calculations using arbitrary
three-dimensional traps. In case of attractive two-body interaction, 
the maximum critical number of atoms for a stable system is one of the 
interesting observables to study, which is also 
related to the collapse of the wave-function of the system. In these cases, 
where the two-body scattering length is negative and the kinetic energy 
cannot be considered to be a small perturbation, the Gross-Pitaevskii mean 
field approximation has been applied, given reliable results in explaining 
the observations in the stable (non-collapsing) conditions~\cite{ketterle}.

Before presenting the mean-field equation for an arbitrary 3D case, 
let us analyze qualitatively the collapse phenomenon for asymmetric traps.
The interaction energy is proportional to the square of the density, varying 
with the negative two-body scattering length. 
For traps with cylindrical (or almost cylindrical) shapes, there are two quite 
different situations: 
one, pancake-like, with the frequencies in the transverse directions being
smaller than the frequency in the longitudinal direction; the other,
cigar-like (quasi-1D), with the frequency in the longitudinal direction 
smaller than the frequencies in the perpendicular directions.
For a true 1D system, one does not expect the collapse of the system with 
increasing number of atoms~\cite{rup95,ND}. However, it happens that 
a realistic 1D limit is not a true 1D system, with the density of 
particles still increasing due to the strong restoring forces in the 
perpendicular directions~\cite{perez98}. 
The relevance of the quasi-1D trap have been pointed out in  
Ref.~\cite{perez98}, to control the condensate motion.
But, as we are going to see, the critical number of particles in the 
quasi-1D limit is smaller than the critical number of particles in the 
2D limit, if we just exchange the longitudinal and perpendicular frequencies.
 The physical reason for that behavior is the increase of the
 average density in the cigar-like configuration relative to the pancake
 like one for the same number of atoms, implying in a strong collapsing 
force in the first case and consequently the cigar-like geometry is a more 
unstable configuration compared to the pancake like one.
This conclusion is in apparent contradiction with
the remark made in section IV of Ref.~\cite{perez98}, saying that, 
considering the better collapse-avoiding 
properties, ``the cigar-shaped trap is the optimal one".  
We are going to discuss this problem in detail and clarify this issue.

In the following, we revise the Gross-Pitaevskii (GP) formalism, for an
atomic system with arbitrary non-spherically symmetric harmonic trap. 
The Bose-Einstein condensate, at zero temperature, in the GP mean-field 
approximation is given by
\begin{eqnarray}
{\rm i}\hbar\frac{\partial}{\partial t}\Psi
(\vec{r},t) 
&=& \left[ -\frac{\hbar ^{2}}{2m}{\bf \nabla }^{2}+
\frac{m}{2}
\left(
\omega_{1}^{2}r_1^{2}+
\omega_{2}^{2}r_2^{2}+
\omega_{3}^{2}r_3^{2}
\right)
\right.\nonumber\\
&+& \left.
\frac{4\pi \hbar ^{2}\ a}{m}|\Psi 
(\vec{r},t)
|^{2}\right]\Psi
(\vec{r},t) 
= \mu\Psi
(\vec{r},t),
\label{sch}
\end{eqnarray}
where $r=\sqrt{r_1^2+r_2^2+r_3^2}$, $m$ is the mass of the atom, $\mu$ 
is the chemical potential, and the 
wave-function $\Psi\equiv\Psi(\vec{r},t)=\Psi(\vec{r},0)
{\rm exp}(-{\rm i}\mu t/\hbar)$ 
is normalized to the number of particles $N$.
The arbitrary geometry of the trap is parameterized by  
three different frequencies  $\omega_1$, $\omega_2$ and $\omega_3$.
For convenience, 
it is appropriate to define the frequencies according to their 
magnitude, such that, in the present work we assume  
$\omega_1\le\omega_2\le\omega_3$.

Here we will be concerned only with systems that have attractive
two-body interactions [$a=-|a|$, in Eq.~(\ref{sch})]. In this case,
it was first shown numerically, in Ref.~\cite{rup95}, that the 
system becomes unstable if a maximum critical number of atoms, $N_c$, is 
achieved. We present precise results  for a critical parameter 
$k$, directly related to the maximum number of atoms, in a general 
non-symmetric configuration of the trap.  

By rewriting Eq.(\ref{sch}) in dimensionless units:
\begin{equation}
{\rm i}\frac{\partial \phi}{\partial\tau} = 
\left[\frac{1}{2}\sum_i\left(-\frac{\partial^2}{\partial x_i^2}
+ 
\frac{\omega_i^2\;x_i^2}{\;\omega^2}\right)
 - |\phi|^2 
\right]\phi
\label{sch2},\end{equation}
where $\tau\equiv\omega t$,
$r_i\equiv l_0\; x_i$ and
$\phi\equiv l_0\sqrt{4\pi|a|}\Psi$, with
\begin{equation}
\int d^3 x |\phi|^2=4\pi\frac{N|a|}{l_0}
\label{philbd}\end{equation}
The oscillator length $l_0$ is defined in terms of $\omega$, which is taken 
as the geometrical mean value of the frequencies:
\begin{equation}
l_0\equiv\sqrt{\frac{\hbar}{m\omega}},\;\;\;{\rm with}\;\;\;
\omega\equiv\left(\omega_1\omega_2\omega_3\right)^{1/3}. 
\label{l0}
\end{equation}
For strong non-symmetric cases, 
particularly when comparing the two extreme cylindrical-shape 
geometries, $\omega_1\sim\omega_2<<\omega_3$ (pancake-shape) and 
$\omega_1<<\omega_2\sim\omega_3$ (cigar-shape), it
is expected a noticeable difference between the corresponding
critical number of particles. 

We define a parameter $k$, related to the critical number of trapped atoms
$N_c$ as in Ref.~\cite{roberts}:
\begin{equation}
k=\frac{N_c|a|}{l_0}.
\label{k}
\end{equation}
This parameter is a maximum critical limit for stable 
solutions of the dimensionless Eq.~(\ref{sch2}).
It will depend only on the ratio of the frequencies of the trap.
Within the precision given in Ref.~\cite{gammal}, $k_s=0.5746$,
where $k_s$ is $k$ for spherically symmetric traps.
In Ref.~\cite{dalfovo}, the critical number was calculated
for a nonsymmetrical geometry, but in a case that the frequency ratio
is not too far from the unity ($\omega_1/\omega_\perp = 0.72$), giving
a result for the number of atoms almost equal to the spherical one.

In the experiments with $^{85}$Rb~\cite{roberts}, 
they  have considered an almost cylindrical ``cigar-type" symmetry, 
with the three frequencies given by 17.47 Hz, 17.24 Hz and 6.80 Hz. 
With this symmetry, they have obtained 
$k=0.459\pm0.012$ (statistical) $\pm 0.054$ (systematic). 
In Ref.~\cite{GFTbr}, assuming the cylindrical symmetry 
$\omega_1=2\pi\times 6.80$Hz $\omega_2=\omega_3=2\pi\times 17.35$ Hz,
it was obtained $k=0.55$, a value about 4\% lower than 
$k_s$. 

In our numerical approach, the calculation is performed by evolving the
nonlinear equation (\ref{sch2}) through imaginary time~\cite{dalfovo}. The
evolution is performed for an initial value of the normalization
(\ref{philbd}) till the wave function relaxes to the ground state.
The wave function is renormalized after each time step.
The process is repeated systematically for larger values of the 
normalization, until a critical limit is reached. At this critical limit 
the ground state becomes unstable. 
The time evolution is done with a semi-implicit second
order finite difference algorithm. An alternating scheme is used in 
the $x_1$ and $x_2$ direction, with a split step in the $x_3$ direction. This
procedure is done only for $x_i\ge 0$, taking advantage of the reflection
symmetry of the ground state. We consider 100$^3$ grid points and time
step equal to $\Delta\tau=0.001$, verifying that the algorithm is stable
for long time evolution.  As we increment the normalization, approaching
the critical limit, the wave function starts to shrink.
So, in order to maintain the precision,
we introduce an automatic reduction of the grid sizes, $\Delta x_1$,
$\Delta x_2$, and $\Delta x_3$, gauged by the respective root
mean-square-radius in each direction. 

In Fig.~1 we show our main results for the critical constant $k$, 
covering many different geometries. We plot $k/k_s$ (and $k$ 
in the rhs $y-$axis) as a function of $\lambda$, which is defined by
\begin{equation}
\lambda\equiv\frac{\omega_1\omega_3}{\omega_2^2}=
\left(\frac{\omega}{\omega_2}\right)^3,\;\;\;{\rm with}\;\;\;
\omega_3\ge\omega_2\ge\omega_1.
\label{lambda}\end{equation}
 \begin{figure}
 \setlength{\epsfxsize}{1.0\hsize}
\centerline{\epsfbox{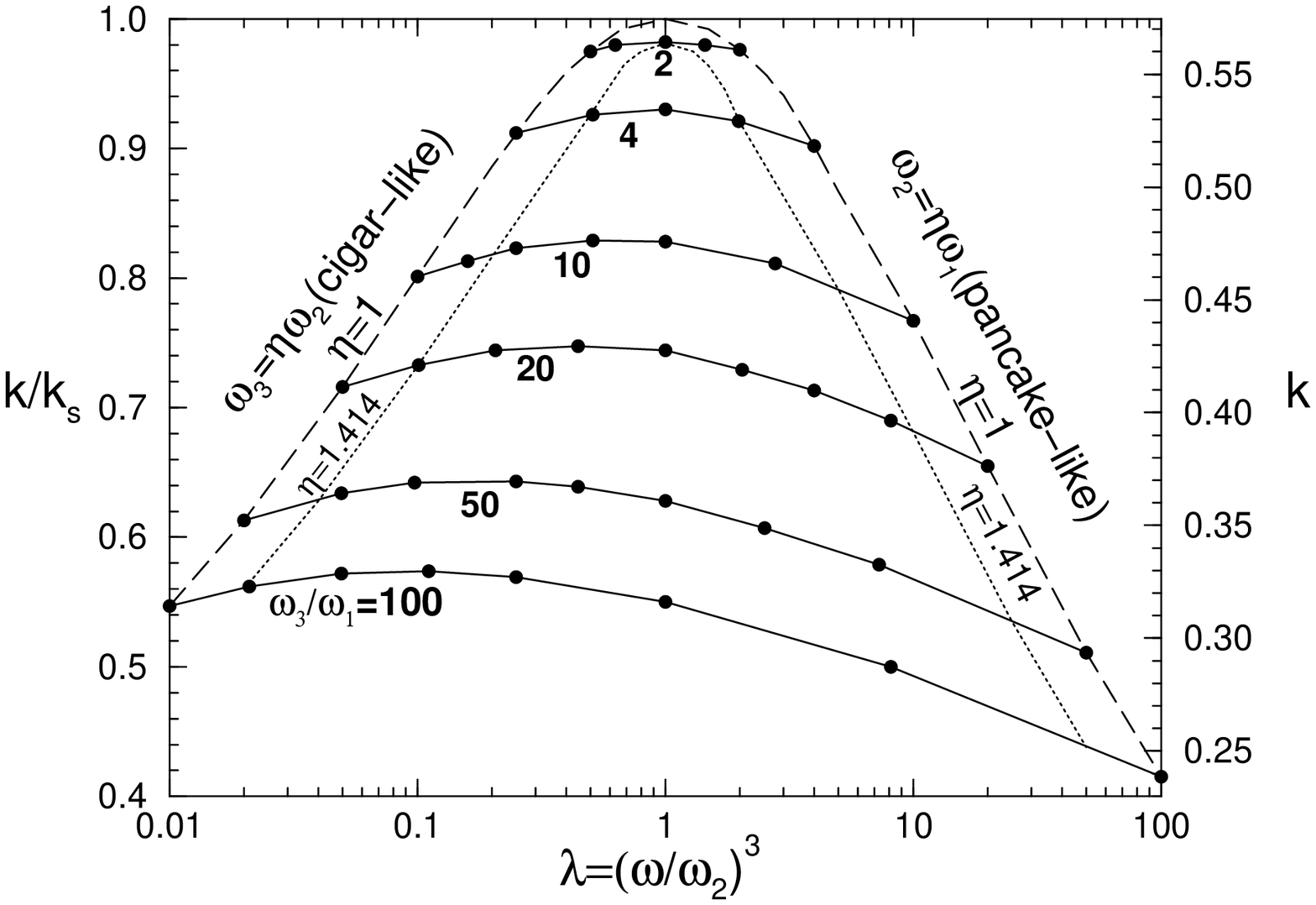}}
 \caption[dummy0]{
Critical constant
$k=N_c|a|(m\omega/\hbar)^{1/2}$, calculated for an arbitrary
non-symmetric trap,
$\omega\equiv(\omega_3\omega_2\omega_1)^{1/3}$, 
with $\omega_3\ge\omega_2\ge\omega_1$.
The ratio $\omega_3/\omega_1$ is shown below each
corresponding curve (solid-lines, with full circles). 
The dashed and dotted-lines correspond to cylindrical ($\eta=1$) 
and deformed cylindrical ($\eta=\sqrt{2}$) symmetries,
respectively.
}
\end{figure}
We show several curves in which we kept constant the ratio 
$\omega_3/\omega_1$  (solid-lines).
The values of $\omega_3/\omega_1$ are indicated inside the 
figure, just below the corresponding plot.
The dashed and dotted-lines correspond to cylindrical ($\eta=1$) and
deformed cylindrical ($\eta=1.414$) symmetries for the trap, where $\eta$ is 
the deformation parameter, as it will be explained. 
In the left-hand-side (lhs) of these plots we have 
$\omega_3=\eta\omega_2>>\omega_1$ (cigar-shape); and, in the right-hand-side, 
$\omega_3>>\omega_2=\eta\omega_1$ (pancake-shape).
$k$ can be determined for any symmetry that was not 
shown, by interpolating the already given results.
The results for the complete non-symmetric case are consistent with the 
previously obtained values in cylindrical symmetry~\cite{GFTbr}.
The maximum value for the critical number $k$ is obtained for the 
spherically symmetric case ($\omega_1=\omega_2=\omega_3$). 

As already observed, Fig. 1 includes previous calculations in the limit of 
the quasi 1D (cigar-shape) and quasi 2D (pancake-shape) symmetries
(see dashed-line). 
However, for the sake of comparison with previous results obtained by 
several authors, we also present in Fig. 2 the cylindrical  
pancake-type ($\omega_\perp=\omega_1=\omega_2<<\omega_3$) and 
cigar-type ($\omega_\perp=\omega_3=\omega_2>>\omega_1$) results.
In Fig. 2, for cylindrical geometries, we compare our exact results, 
for $k_3=N_c|a|(m\omega_3/\hbar)^{1/2}$ as a function of $\lambda$,
with the corresponding variational ones of Refs.\cite{baym,perez97}. 
The variational results (dashed-lines) are consistently a bit higher than 
the exact ones.
For $\lambda\to 0$ ($\omega_1<<\omega_2=\omega_3=\omega_\perp$),  
the exact and variational results for the critical 
constant are $N_c|a|(m\omega_\perp/\hbar)^{1/2}=$0.676 and 0.776, 
respectively. They are consistent with the quasi-1D limits given in 
Refs.~\cite{perez97,perez98}.
When $\omega_3>>\omega_2=\omega_1$, the variational 2D
limit $\sqrt{\pi/2}$ of Ref.~\cite{baym} is comparable with our
exact result ($k_3=0.931\sqrt{\pi/2}$).
{\it In this case, the quasi-2D limit coincide with the 
true 2D limit}~\cite{weinstein,adhikari,ND,berge}. 

 \begin{figure}
 \setlength{\epsfxsize}{0.8\hsize}
\centerline{\epsfbox{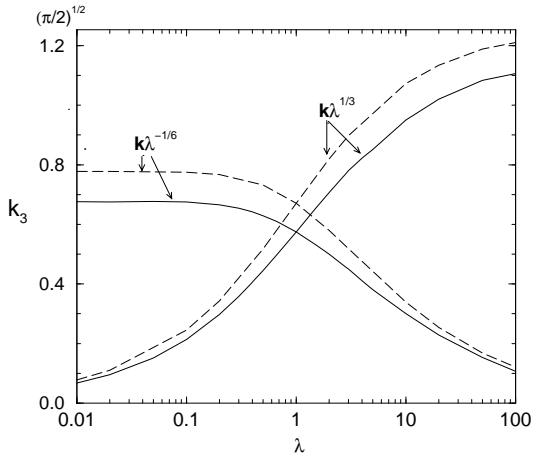}}
 \caption[dummy0]{
Exact results of $k_3\equiv N_c|a|(m\omega_3/\hbar)^{1/2}$,
for cylindrical traps (solid-lines) are compared with
corresponding variational approaches (dashed-lines).  
In the 2D pancake limit, $\omega_\perp=\omega_2=\omega_1$ and
$k_3=k\lambda^{1/3}$;  and, in the quasi 1D cigar limit, 
$\omega_\perp=\omega_2=\omega_3$
and $k_3=k\lambda^{-1/6}$.
Note that $\lambda=\omega_3/\omega_1$, when $\omega_2=\omega_1$; 
and $\omega_1/\omega_3$, when $\omega_2=\omega_3$.
} 
\end{figure}

Considering the present analysis, we observe that $N_c$ 
in a cigar-like (quasi-1D) trap is smaller than $N_c$ in a pancake-like 
(quasi-2D) trap. And, as we deform a cigar-like trap, $N_c$ also increases.
We need to clarify this matter, which is the subject of the next four 
paragraphs, because such result is apparently contradicting a remark made in 
Ref.~\cite{perez98}, saying that ``the cigar-shaped trap is the optimal one". 

As shown in  Ref.~\cite{GFTbr} and also in the present calculation, in
a deformed cylindrical symmetry, the cigar shape (with one 
of the frequencies smaller than the other two) is more favorable to 
obtain a larger value of $k$ than the pancake-shape symmetry (with one
of the frequencies larger than the other two). See, for example, in Fig. 1, 
the two extreme points of the curve with $\omega_3/\omega_1 = 100$.
This results from the definition of $k$, Eq.~(\ref{k}), in terms of the 
averaged oscillator length $l_0$. However, the maximum value of $k$ can 
only be directly related with the maximum value of $N_c$ in case that  
$\omega$ is kept fixed. 
And, with $l_0$ fixed, $N_c$ is maximized for 
$\lambda=1$, corresponding to the spherically symmetric case ($k=k_s$).
If we fix $l_0$ and $\omega_3/\omega_1$, $N_c$ is maximized for a deformed
cylindrical symmetry with $\lambda<1$, as one can see from Fig. 1.

Considering exact cylindrical traps, by exchanging the frequencies (which, 
obviously, does not keep constant the averaged frequency $\omega$), it is 
valid the following ratio that was obtained in Ref.~\cite{GFTbr}: 
\begin{equation}
R(\lambda)\equiv\frac{N_c(\lambda)}{N_c(1/\lambda)}=
\lambda^{1/6}\frac{k(\lambda)}{k(1/\lambda)}.
\label{R}\end{equation}
This result favors the pancake-like symmetry ($\lambda=
\omega_3/\omega_1=\omega_3/\omega_\perp>1$), to obtain a larger value for $N_c$.
Consider, for example, a cylindrical pancake-type trap with $\lambda=100$, in 
comparison with a cigar-type trap with $\lambda=1/100$. We notice that,
in this case, $R(\lambda=100)\approx 1.6$, implying that
with such pancake-like trap ($\omega_3=100\omega_\perp$)
one can obtain about 60\% more particles than with the corresponding 
cigar-like trap ($\omega_3=\omega_\perp=100\omega_1$).
Let us consider the recent experiment with quasi-1D 
(cigar-like) trap used in the formation and propagation of matter 
wave solitons, with $^7$Li~\cite{strecker}. In this case, it was 
used axial and radial frequencies, respectively, equal to
3.2 Hz and $\sim$625 Hz~\cite{private}; or, 
$\omega_\perp=\omega_3=2\pi\times$ 625 Hz, 
$\omega_1=2\pi\times$ 3.2 Hz and 
$\lambda=\omega_1/\omega_3=$ 0.00513. 
So, as shown in Fig.2, we are practically in the limit $\lambda=0$,
which gives $N_c^{1D}|a|/\sqrt{\hbar/(m\omega_\perp)}\approx 0.675$. 
Considering that the scattering length was tuned to $a=-3 a_0$ ($a_0$ is 
the Bohr radius), 
the maximum number of atoms in this quasi-1D trap is 
$N_c^{1D}\approx 6400$. If we exchange the radial and axial 
frequencies in this experiment, going from a cigar-like to 
a pancake-like trap, 
$\omega_\perp=2\pi\times$ 3.2 Hz, $\omega_3=2\pi\times$ 625 Hz and 
$\lambda=\omega_3/\omega_\perp= 195.3$. In this case,
$N_c^{2D}|a|/\sqrt{\hbar/(m\omega_3)}\approx 1.12$. 
So, the critical number in the quasi-2D limit is about 66\% larger 
than the corresponding number of atoms in the quasi-1D limit, and 
$N_c^{2D}\approx 10600$.

The discussion about the best way to distribute the frequencies to obtain
the maximum number of atoms was first considered in Ref. \cite{sackett}, 
arriving that the best is to do a spherical trap, for achieving maximum 
density.
(In their almost spherical trap they have obtained from $\sim$600 to 
$\sim$1300 atoms, in an overall agreement with theoretical 
predictions {\bf \cite{hulet95}}.)
The point is that, to say what is the best configuration for the maximum 
number of atoms we must also say what is the {\it constraint} in the 
frequencies. If the constraint is to keep fixed the product 
$\omega_1 \omega_2 \omega_3$ than the maximum number will be given for the 
maximum $k$ [as shown in Eq.(\ref{k})], that happens in a spherical symmetry, 
in agreement with Ref.\cite{sackett}.  
But, if two frequencies are equal and given {\it a priori}, than the best 
configuration to increase $N_c$ is to make the third frequency equal to zero, 
in agreement with Ref.~\cite{perez98}.
However, if only one frequency is kept fixed, than the extreme pancake shape 
will contain more atoms, by making the remaining frequencies go to zero.

As a final remark about the maximization of $N_c$, we note that, 
for attractive two-body interactions, 
the frequencies in all three directions cannot be initially zero. 
In the experimental process of condensation, the atoms must be trapped 
while going from higher temperature $T$ to the critical one, $T_c$.
By making the frequencies too small, the $T_c$ will be smaller and the 
condensation harder to achieve [as shown in Ref.~\cite{DGP}, $T_c\sim 
\hbar\omega(N_{tot})^{1/3}$, where $N_{tot}$ is the total number of particles 
in the trap].
Once the condensate is achieved, we can switch off some of the frequencies,
and obtain a finite maximum value for $N_c$, as observed in Fig. 2 for $T=0$.
In the present work, we are concerned with the case of zero temperature. 
Effects of temperature on the collapse and on $N_c$, for spherically 
symmetric case, have been discussed, for example, in Ref.~\cite{DHZ}.
$N_c$ is maximized by making $\omega$ as small as possible. 
However, a discontinuity exists when taking all the three frequencies 
exactly zero. Without trap the collapse will occur for any number of particles.

We must stress that Fig. 1 also includes novel non-symmetric cigar-type 
geometries $\omega_3=\eta\omega_2 >> \omega_1$ and elliptical pancake-type 
geometries $\omega_3>>\omega_2=\eta\omega_1$, where $\eta$ is the excentricity 
of the ellipsis. 
In particular, the dotted line in Fig.1 gives the
critical values of $k$ for $\eta=\sqrt{2}$.
Thus, one could replot, as in Fig. 2, a series of curves whose limiting 
cases describe non-symmetric cigar- or pancake-like symmetries. In the 1D 
and 2D limits, one can obtain critical quasi-1D and quasi-2D numbers, for 
each $\eta$.

One can also obtain an interesting result, for asymmetric cigar-shape traps, 
where $\omega_3=\eta \omega_2>> \omega_1$, following
the approach given in section II of Ref.~\cite{perez98}.
The assumption made in \cite{perez98} to reduce the 3D equation to a 1D 
equation can be justified in the limit when the forces in the transverse 
directions are the main responsible for the trapping potential.
So, in the limit $\omega_1\to 0$, we are able to generalize the
solitonic 1D equation for a cigar-shaped trap, deformed by a given
ratio $\eta=\omega_3/\omega_2$ between the transverse direction frequencies.
In this case, $\eta\ge 1$ cannot be arbitrarily larger.
Also, by a scaling procedure applied to the equation for different values of 
$\eta$, using our previous exact result, we have
\begin{equation}
{N_c|a|}\left(\frac{m\sqrt{\omega_2\omega_3}}{\hbar}\right)^{1/2}=
\frac{N_c|a|}{\eta^{1/4}}\sqrt{\frac{m\omega_3}{\hbar}}=0.676
.\label{Ncigar}\end{equation}
This generalizes the cigar-shape quasi-1D results of \cite{perez98} (with 
$\eta$ such that we still have $\omega_2>>\omega_1$).  
The 1D solid line given in Fig.2 is also applied to deformed 
cigar-shaped symmetries if we replace $k_3$ by $k_3 \eta^{-1/4}$ in the 
$y-$axis.
This result may be relevant for asymmetric waveguide propagation as one 
can deform the cigar-type symmetry and control the collapsing condition. 
From Eq.~(\ref{Ncigar}) we observe that the maximum critical number $N_c$ 
will increase when deforming the cigar-like symmetry by a factor 
proportional to $(\omega_3/\omega_2)^{1/4}$.

Now, let us see the effect of deformation in the same example 
of the cigar-shaped geometry used in Ref.~\cite{strecker}, with
$^7$Li gas. We use the same value of $a=-3a_0$, with
$\omega_3=2\pi\times$ 625 s$^{-1}$, and take
$\omega_2=\omega_3/\eta$. As given in Eq.~(\ref{Ncigar}), we
go from $N_c=6392$ $(\eta=1)$ to $N_c=6970$ $(\eta=\sqrt{2})$,
or $N_c=7601$ $(\eta=2)$. It means, an increase of 
$\sim$9\% when $\eta=\sqrt{2}$; and $\sim$19\% when $\eta=2$.
In case, $\eta>>1$, the approximation considered in the wave-function
separation, as given in \cite{perez98}, is not valid. In such
case we are reaching the other deformed pancake-like symmetry, where
$\omega_3>>\omega_2\sim\omega_1$. However, in the pancake geometry, 
the effect of deformation in $N_c$ is negligible. 
By comparing the quasi pancake-like geometry with the
quasi cigar-like geometry, the number $N_c$ in the cigar-like geometry 
is much more sensitive to deformations. 

 \begin{figure}
 \setlength{\epsfxsize}{1.0\hsize}
\centerline{\epsfbox{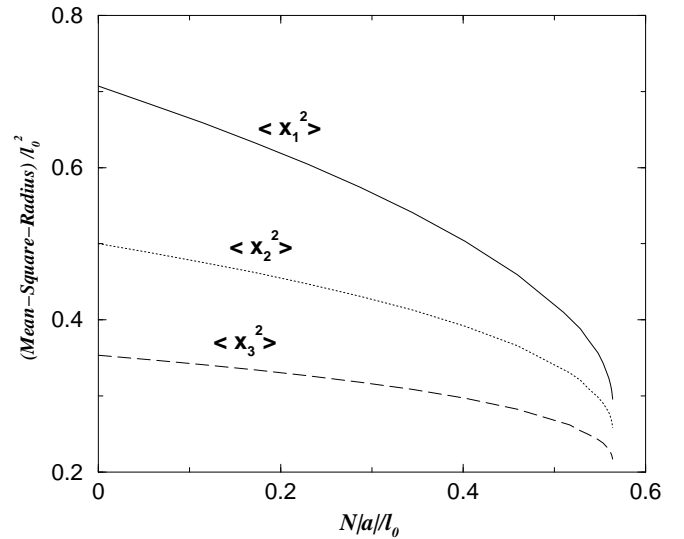}} 
 \caption[dummy0]{The three components, $i=1,2,3$, of the 
mean-square-radius, $\langle x^2_{i}\rangle=\langle r^2_{i}\rangle/l_0^2$, 
are shown in terms of the dimensionless $N|a|/l_0$,
for the case that $\omega_3=\sqrt 2\;\omega$, $\omega_2=\omega$, and 
$\omega_1=\omega/\sqrt 2\;$. }
\end{figure}

We have also studied the behavior of the root mean square radius for 
the case $\omega_3=\sqrt{2}\omega_2=2\omega_1$.  
We verified that, as the system approaches the critical point (or 
collapse), the wave function tends to be more ``spherical", confirming 
earlier conclusion made with a Gaussian variational 
approximation~\cite{pires}. 
In Fig. 3, we show the corresponding results, for the three 
components of the mean-square-radius, 
$\langle r^2_{i}\rangle$ $= l_0^2\langle x^2_{i}\rangle\; (i=1,2,3)$.
As shown, when $N=0$, we have $\langle r_1^2\rangle/\langle r_2^2\rangle=$
$\langle r_2^2\rangle/\langle r_3^2\rangle=\sqrt 2$; and,
when $N\approx N_c$ such ratio is drastically reduced.

In summary, we have calculated systematically
the critical number of particles, in systems that have negative two-body 
interactions, for traps with arbitrary geometries. 
The maximum critical number of particles $N_c$ can be derived from 
the given value of the parameter $k$, given in Eq.~(\ref{k}), once one has 
the scattering length, the mass of the atomic system, and the 
frequencies of the trap. The results are shown in Figs. 1 and 2.
The value of $k$, for any symmetry non explicitly given,
can be easily derived from Fig. 1, by interpolation.
It is also pointed out that the results shown in Fig. 2,
in the 1D cigar-like case, can be extended for slightly
deformed cylindrical symmetries, by replacing the $y-$axis label $k_3$ by 
$k_3\eta^{-1/4}$.

Our main results in the present work are:
(i) The maximum number of particles, $N_c$, for arbitrary 3D trap geometries 
is given through the results shown in Fig.1.
(ii) The optimal trap configuration, to avoid the collapse with maximum $N_c$, 
is found to be strongly dependent on the constraints of the frequencies of the 
trap. 
If we initially fix one of the frequencies, the best configuration of the trap 
is pancake-like, with the other two frequencies going to zero.
Analogously, if we initially fix two equal frequencies, the best configuration 
of the trap is cigar-like, with the third frequency close to zero.
If we initially fix two different frequencies and try to vary the third 
frequency between the fixed ones, the best configuration is again pancake-like.
We show that $N_c$ is much more sensitive to deformations of the trap in a 
cigar-like geometry than in a pancake-like geometry. 
Finally, for small deformations $\eta$ of the cigar-like traps, where 
$\eta\ge 1$, releasing the longitudinal direction, 
the solitonic solutions, obtained in
Ref.~\cite{perez98} will be rescaled by the deformation. 
$N_c$ will be rescaled by a factor $\eta^{1/4}$, generalizing the
findings of Ref.~\cite{perez98}.

\section*{Acknowledgements}
We are grateful to Randy Hulet for informations provided.
AG also thanks Emerson Passos and Marcelo Pires for useful 
discussions. For partial support, we thank Funda\c c\~ao de Amparo \`a 
Pesquisa do Estado de S\~ao Paulo (FAPESP) and Conselho Nacional de 
Desenvolvimento Cient\'\i fico e Tecnol\'ogico (CNPq).

\end{multicols}
\end{document}